\documentclass[conference, compsocconf]{IEEEtran}
\IEEEoverridecommandlockouts
\usepackage{cite}
\usepackage{listings}
\usepackage{amssymb}
\usepackage{bm}
\usepackage{amsmath,amssymb,amsfonts}
\usepackage{algorithmic}
\usepackage{graphicx}
\usepackage{enumerate}
\usepackage{textcomp}
\usepackage{xcolor}
\usepackage{multirow}
\usepackage{url}
\def\BibTeX{{\rm B\kern-.05em{\sc i\kern-.025em b}\kern-.08em
    T\kern-.1667em\lower.7ex\hbox{E}\kern-.125emX}}

\lstdefinelanguage{md}{
  sensitive=true,
  showstringspaces=false,
  moredelim=[s][\bfseries]{__}{__},
  moredelim=[s][\ttfamily\color{orange!90!black}]{`}{`},
  moredelim=[s][\color{orange!90!black}]{**}{**},
  morecomment=[l][\bfseries]{\#},
  morecomment=[l][\bfseries]{\#\#},
  morecomment=[l][\color{violet}\bfseries]{\#\#\#},
  morecomment=[l][\color{violet}]{\#\#\#\#},
  morecomment=[l][\color{violet}]{\#\#\#\#\#},
}

\lstdefinestyle{mdstyle}{
  language=md,
  basicstyle=\ttfamily\scriptsize,
  breaklines=true,
  columns=fullflexible,
  keepspaces=true,
  upquote=true,
  frame=tbrl,
  rulecolor=\color{black},
  framesep=4pt,
  xleftmargin=8pt,
  xrightmargin=8pt,
  captionpos=t,
  literate={\\_}{\_}1,
}

\usepackage{caption}
\begin{document}

\title{VibeCodeHPC: An Agent-Based Iterative Prompting Auto-Tuner for HPC Code Generation Using LLMs}

\makeatletter
\newcommand{\linebreakand}{%
  \end{@IEEEauthorhalign}
  \hfill\mbox{}\par
  \mbox{}\hfill\begin{@IEEEauthorhalign}
}
\makeatother

\author{
\IEEEauthorblockN{Shun-ichiro Hayashi}
\IEEEauthorblockA{Graduate School of Informatics\\
Nagoya University\\
Aichi, Japan\\
hayashi@hpc.itc.nagoya-u.ac.jp}
\and
\IEEEauthorblockN{Koki Morita}
\IEEEauthorblockA{Graduate School of Informatics\\
Nagoya University\\
Aichi, Japan\\
morita@hpc.itc.nagoya-u.ac.jp}
\and
\IEEEauthorblockN{Daichi Mukunoki}
\IEEEauthorblockA{Information Technology Center\\
Nagoya University\\
Aichi, Japan\\
mukunoki@cc.nagoya-u.ac.jp}
\linebreakand
\IEEEauthorblockN{Tetsuya Hoshino}
\IEEEauthorblockA{Information Technology Center\\
Nagoya University\\
Aichi, Japan\\
hoshino@cc.nagoya-u.ac.jp}
\and
\IEEEauthorblockN{Takahiro Katagiri}
\IEEEauthorblockA{Information Technology Center\\
Nagoya University\\
Aichi, Japan\\
katagiri@cc.nagoya-u.ac.jp}
}

\maketitle

\begin{abstract}
In this study, we propose VibeCodeHPC, a multi-agent system based on large language models (LLMs) for the automatic tuning of high-performance computing (HPC) programs on supercomputers. VibeCodeHPC adopts Claude Code as its backend and provides an integrated environment that facilitates program development in supercomputer settings. The system not only brings the Vibe Coding paradigm -- program development through natural language interaction with users -- to HPC programming, but also enables autonomous performance optimization with minimal user intervention through a sophisticated multi-agent design. To achieve these objectives, VibeCodeHPC implements three core functionalities: (1) configuration capabilities tailored to the unique development environments of supercomputers, (2) collaborative operation among multiple LLM agents with distinct roles -- Project Manager (PM), System Engineer (SE), Programmer (PG), and Continuous Deliverer (CD), and (3) long-term autonomous operation through agent activity monitoring and dynamic deployment mechanisms. This paper highlights one of the most powerful features of VibeCodeHPC: fully automated code optimization through autonomous operation without user intervention. Specifically, it demonstrates the performance optimization of CPU-based codes on GPU-equipped systems for matrix multiplication and a Poisson equation solver using Jacobi's iterative method. The results show that the multi-agent configuration employed in VibeCodeHPC enables faster and more reliable development of higher-performance code compared to a single-agent setup.
\end{abstract}

\begin{IEEEkeywords}
Code Generative AI, Auto-tuning, Vibe Coding, Multi AI Agents, Software Productivity
\end{IEEEkeywords}

\section{Introduction}
In recent years, large language models (LLMs) have shown strong capabilities in automatic code generation, transforming natural language specifications into executable programs. Trained on large-scale source code corpora, these models can generate syntactically correct and semantically meaningful code and have been applied to software development tasks such as prototyping, bug fixing, refactoring, and testing. Prior work has demonstrated that large-scale code-trained models can generate effective solutions for programming problems \cite{chen2021evaluating}, and that prompt engineering and in-context learning improve code correctness and robustness \cite{austin2021program}\cite{li2022alphacode}.
More recently, coding agent systems have emerged that autonomously perform user-specified tasks through iterative execution and refinement. Representative examples include Claude Code\footnote{\url{https://github.com/anthropics/claude-code}}, Codex CLI\footnote{\url{https://github.com/openai/codex}}, and Gemini CLI\footnote{\url{https://github.com/google-gemini/gemini-cli}}, which integrate LLMs with command-line interfaces (CLI). However, despite these advances, existing LLM-based systems primarily target general-purpose programming and do not sufficiently address performance-critical domains such as high-performance computing (HPC) applications on supercomputers, particularly with respect to performance optimization.
\par
HPC programming imposes stringent requirements beyond functional correctness, including architecture-specific optimizations, parallelization across threads and processes, and trade-offs between numerical precision, performance, and power consumption. HPC applications typically rely on parallel programming models such as MPI, OpenMP, CUDA, and OpenACC, which must be selected according to application characteristics and execution environments. In addition, supercomputer usage requires system-specific operations, including batch job execution, software module management, and Secure Shell (SSH) access.
\par
These complexities pose significant challenges for both human developers and code generation AI agents. LLMs, like humans, are constrained by limited context windows, and prolonged complex tasks may lead to forgotten instructions or degraded behavior. Multi-agent systems, in which multiple LLMs with specialized roles collaborate, offer a promising approach to mitigating these limitations by distributing cognitive load across agents. This advantage is particularly relevant for complex domains such as HPC programming. Moreover, defining rigorous requirements in advance is often difficult in such settings. To address this, the Vibe Coding paradigm has been proposed \cite{karpathy2025vibecoding}\cite{meske2025vibecoding}, in which programs are incrementally developed through interactive dialogue between humans and LLMs without strict upfront specifications. This paradigm is well suited to complex development domains, including HPC programming.
\par

Based on the above background, this study proposes VibeCodeHPC, a multi-agent system based on LLMs for the automatic tuning of HPC programs on supercomputers. VibeCodeHPC is built on top of Claude Code and offers an integrated framework for program development targeting supercomputing environments. Thanks to its sophisticated multi-agent mechanism, the system not only enables vibe coding as if working with a ``specialized HPC programming team,'' but also achieves fully automated tuning through autonomous operation that minimizes user intervention. To realize these objectives, VibeCodeHPC provides the following core functionalities:
\begin{enumerate}
    \item Configuration capabilities tailored to the unique development environments of supercomputers
    \item Collaborative operation among multiple LLM agents with distinct roles -- Project Manager (PM), System Engineer (SE), Programmer (PG), and Continuous Deliverer (CD)
    \item Long-term autonomous operation through agent activity monitoring and dynamic deployment mechanisms
\end{enumerate}
\par

This paper is organized as follows. Section~\ref{sec:related} reviews related work. Section~\ref{sec:overview} presents VibeCodeHPC and its implementation, focusing on multi-agent features. Section~\ref{sec:eval} demonstrates automated code optimization through autonomous operation. Finally, Section~\ref{sec:conclusion} summarizes the findings and discusses future research directions.
\par 

\section{Related Work}
\label{sec:related}

\subsection{HPC Code Generation and Optimization}
HPC code generation has traditionally relied on domain experts to manually implement performance-critical kernels using parallel programming models such as MPI, OpenMP, CUDA, and OpenACC. Compiler-assisted approaches \cite{wolfe1995compilers}\cite{kennedy2001optimizing} partially automate this process through loop transformations, memory layout optimization, and vectorization, while domain-specific languages (DSLs) \cite{vasilache2018tensor}\cite{kamil2010autotuning} generate optimized kernels from high-level specifications. Although these methods can achieve high performance, they often require explicit user annotations or DSL expertise, limiting accessibility.
\par 

Recent studies have begun evaluating LLMs for parallel programming models and HPC kernels, including OpenMP-centric models and MPI+OpenMP code generation in scalable workloads \cite{ompgpt2024} \cite{llmhpcpp2025}. Early evaluations of LLMs for OpenMP/MPI/CUDA/OpenACC kernels \cite{codexhpc2023} \cite{llama2hpc2023} and LLM-assisted portability from Fortran to C/C++ \cite{fortranport2025} suggest that HPC-specific constraints substantially affect correctness and performance outcomes. These works provide concrete baselines against which VibeCodeHPC can be positioned.
\par 

Practical case studies that combine LLM-based coding agents with GPU acceleration and testing pipelines offer direct evidence of feasibility in HPC contexts \cite{geofem2026}. Additionally, HPC-oriented LLM systems and evaluations for software development provide a broader landscape for integration and operational support \cite{chathpc2025, hpcgpt2023, hpcelm2024, openmpassess2024}.
\par 

\subsection{Auto-tuning and Iterative Refinement}
Auto-tuning (AT) has been widely adopted in HPC to explore parameter spaces and identify optimal configurations for target architectures. Frameworks such as ATLAS \cite{whaley2001atlas}, OpenTuner \cite{ansel2014opentuner}, and GPTune \cite{liu2021gptune} automate this process using search algorithms including Bayesian optimization, genetic algorithms, and multi-armed bandits. In addition, iterative refinement strategies, in which code is repeatedly modified based on performance feedback, have been studied in machine learning–guided compilation \cite{cummins2017deeplearning} \cite{hajali2020neurovectorizer}. While these approaches effectively optimize existing code, their integration with LLM-driven code synthesis remains limited. Comparative studies contrasting agentic LLM approaches with classical autotuning for loop transformations indicate that LLM-driven methods can complement or accelerate parameter search in certain settings \cite{agenticautotune2025}. These findings motivate the combination of structured tuning with multi-agent orchestration.
\par 

\subsection{Iterative Prompt in LLM-based Code Generation}
Recent studies have explored iterative prompting in code generation AI. DiCuffa et al. \cite{dicuffa2025prompt} used the DevGPT dataset to demonstrate that structural prompt patterns enable high-quality outputs with fewer iterations. Sorokin et al. \cite{sorokin2025iterative} introduced an iterative self-learning loop with re-ranking models and approximate policy optimization, achieving performance surpassing GPT-4. Ye et al. \cite{ye2025alchemy} proposed Prochemy, which automatically refines initial prompts based on execution results, enhancing performance in code translation and HumanEval tasks.
\par 

However, iteration can introduce vulnerabilities. Shukla et al. \cite{shukla2025security} reported a 37.6\% increase in critical vulnerabilities after five iterations, highlighting the need for human oversight. From a structural perspective, Ren et al. \cite{ren2023kpc} proposed Knowledge-driven Prompt Chaining (KPC) for exception handling, while Li et al. \cite{li2025structured} introduced Structured CoT, improving Pass@1 accuracy by up to 13.79\% on HumanEval. Ridnik et al. \cite{ridnik2024alphacodium} demonstrated that test-driven, multi-stage iterative flows improved Pass@5 accuracy from 19\% to 44\%. Wei et al. \cite{wei2025astra} investigated CUDA code optimization using multi-agent systems without additional training. However, the role-sharing approach proposed in our study has not been evaluated for cross-language optimization. These studies constitute methodological investigations into iterative prompting. Research on AI-agent collaboration paradigms, as exemplified by VibeCodeHPC, remains nascent.

\subsection{Multi-Agent Frameworks and Orchestration Tools}
Beyond research papers, the recent surge of general-purpose multi-agent frameworks (e.g., AutoGen\cite{autogen2023}, MetaGPT\cite{metagpt2023}, CAMEL\cite{camel2023}, and Magentic-One\cite{magenticone2024}) and agent orchestration tools (Claude Code subagents\footnote{\url{https://docs.anthropic.com/en/docs/claude-code/sub-agents}}, SuperClaude\footnote{\url{https://github.com/NomenAK/SuperClaude}}, Claude-Flow\footnote{\url{https://github.com/ruvnet/claude-flow}}, Claude Squad\footnote{\url{https://github.com/smtg-ai/claude-squad}}, Vibe Kanban\footnote{\url{https://github.com/BloopAI/vibe-kanban}}, KAMUI\footnote{\url{https://3rd.kamui.ai/en}}) highlights the growing interest in collaborative LLM workflows. Microsoft's agent orchestration patterns categorize \emph{Magentic orchestration} as a distinct mode, which aligns closely with VibeCodeHPC's collaborative, role-separated workflow; however, a framework specialized for supercomputer code development remains unproposed\cite{msmagentic2025}.
\par 

\par 


\subsection{Multi-agent for HPC Code Generation}
Recent HPC-focused multi-agent systems have begun to target code generation, testing, and optimization workflows. Examples include multi-agent HPC code generation and optimization \cite{marco2025}, MPI code synthesis with LLMs \cite{chatmpi2026}, and multi-agent unit test generation for HPC applications \cite{hpcagenttester2025}. Profiling-guided autonomous code generation for parallel workloads is also emerging as a closely related direction \cite{paracodex2026}. These approaches highlight the growing feasibility of collaborative agentic workflows in HPC and motivate VibeCodeHPC's emphasis on role separation, monitoring, and dynamic agent deployment.
\par 
\subsection{Summary}
Existing research in LLM-based code generation has achieved progress in producing functionally correct programs, but it rarely addresses HPC-specific performance optimization. On the other hand, the HPC community has developed powerful auto-tuning and compiler-based optimization techniques, yet these methods often lack the flexibility of LLM-based code synthesis. Bridging these domains -- by integrating LLM-driven generation, HPC performance tuning, and iterative improvement—remains an open challenge, which our proposed VibeCodeHPC framework seeks to address.
\par 

\begin{figure*}[t]
    \centering
    \includegraphics[width=0.8\linewidth]{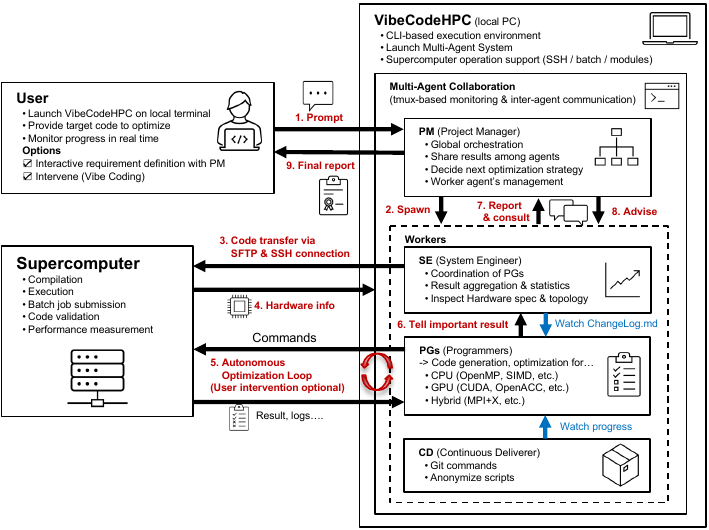}
    \caption{Workflow of VibeCodeHPC.}
    \label{fig:overview}
\end{figure*}

\section{Overview and Implementation of VibeCodeHPC}
\label{sec:overview}
This section explains the system overview of VibeCodeHPC, the operational flow from the user's perspective, and the implementation realizing it. In particular, we provide explanation of the multi-agent implementation, which is the core innovation of VibeCodeHPC.
\par 

\subsection{System Components}
VibeCodeHPC consists of the following main components.
\par 

\begin{itemize}
\item \textbf{CLI-based Coding AI Agent}: This component is the LLM agent that actually performs autonomous programming. The current VibeCodeHPC adopts Claude Code, but conceptually it is also possible to adopt other CLI coding agents. Claude Code is an LLM-based coding assistance tool developed by Anthropic that operates in CLI environments. It receives instructions from users and has the ability to autonomously perform code generation, editing, execution, debugging, and other tasks.
\item \textbf{Functionalities for Supercomputer Operation and Development}: CLI coding agents such as Claude Code are designed with general-purpose programming applications in mind. Therefore, additional support is required when handling operations and functions specific to supercomputers. Accordingly, VibeCodeHPC provides the following supercomputer-specific functionalities:
\begin{itemize}
\item SSH connection management: Establishment and maintenance of remote access to supercomputers
\item Batch job system operation: Job submission, execution status monitoring, and result retrieval
\item Software module management: Loading compilers and libraries using the environment module command
\item Hardware information provision: System information such as cluster configuration and GPU availability
\item Software environment information provision: Information about available compilers, libraries, etc.
\end{itemize}
This information is provided to support agents in appropriately performing system-specific operations.
\item \textbf{Multi-agent Collaborative Operation Functionality}: This is the core functionality provided by VibeCodeHPC. Since coding agents such as Claude Code operate as single-agent systems by default, various implementations are necessary to enable collaborative operation of multiple agents. This component includes the following functionalities:
\begin{itemize}
\item Agent configuration and role setting: Definition of each agent's responsibility scope and operational rules
\item Inter-agent communication mechanism: Message transmission and reception between agents
\item Multi-agent operation rules: Control rules for collaborative operation
\item Agent operation monitoring functionality: Real-time monitoring of each agent's activity status
\item Dynamic agent deployment functionality: Addition and removal of agents as needed
\end{itemize}
Details of this component are described in Section~\ref{sec:multiagent}.
\end{itemize}
\par 


\subsection{Workflow}
Figure~\ref{fig:overview} illustrates the workflow of VibeCodeHPC, which is described as follows.
\par 

\begin{enumerate}
\item \textbf{System Startup}: Although VibeCodeHPC can be operated anywhere, considering the usage pattern of supercomputers, it is assumed to run on the user's local Linux terminal that accesses the supercomputer, from which it launches Claude Code.
\item \textbf{Provision of Tasks and Requirement Definitions}: Users provide VibeCodeHPC with the code they want to optimize for performance and, if necessary, a requirement definition document. When utilizing the Vibe Coding development paradigm, a requirement definition document is not mandatory, and users can proceed with work incrementally while interacting with the system. However, when autonomous code optimization without human intervention is desired, it is recommended to prepare the requirement definition document as rigorously as possible. VibeCodeHPC provides a template for requirement definition documents targeting general supercomputers to support user creation. Items to be described in the requirement definition document include:
\begin{itemize}
\item Target hardware: CPU architecture, GPU availability, node configuration, etc.
\item Performance goals: Target execution time, throughput, energy efficiency, etc.
\item Task constraints: Parts prohibited from optimization, prohibited libraries, precision requirements, etc.
\end{itemize}
The more specific these descriptions are, the higher the likelihood of improved LLM work accuracy.
\item \textbf{Execution of Autonomous Optimization}: When a requirement definition document is provided, VibeCodeHPC begins the optimization process. The code generation\footnote{The code generation LLM itself runs on Anthropic’s cloud.} is performed on the local side, and the generated code is then transferred to the supercomputer via the Secure File Transfer Protocol (SFTP). Subsequently, compilation and execution are carried out on the supercomputer through SSH. For efficient management of SSH and SFTP sessions, VibeCodeHPC employs Desktop Commander MCP\footnote{https://github.com/wonderwhy-er/DesktopCommanderMCP}, which is particularly recommended for supercomputers that require two-factor authentication. Job execution is managed by batch job schedulers such as PBS Pro or Slurm. The entire workflow is orchestrated by Claude Code, while VibeCodeHPC provides the corresponding operational instructions. VibeCodeHPC operates continuously until the user-specified goal is achieved, but users can intervene at any time if necessary. When adopting the Vibe Coding development paradigm, users can appropriately interact with VibeCodeHPC and incrementally refine the optimization direction.
\item \textbf{Monitoring and Result Confirmation}: Users can monitor each agent’s operation in real time, and VibeCodeHPC automatically generates work logs that allow users to track optimization progress. When the user’s objectives are achieved, a final report is generated, summarizing the optimization process, achieved performance, and applied techniques.
\end{enumerate}
\par 

\subsection{Multi-agent Mechanism}
\label{sec:multiagent}
For multiple agents to operate collaboratively and perform productive activities as a multi-agent system, various technical challenges must be overcome. Particularly in programming for supercomputers, problems arise that cannot be addressed by general multi-agent systems due to the complexity of their operation and programming. VibeCodeHPC implements various innovations to overcome these challenges.
\par 

\subsubsection{Agent Configuration}
VibeCodeHPC enables collaborative operation of multiple agents specialized for different tasks. Their roles are defined in markdown-format configuration files loaded by Claude Code at startup and can be customized.
\par 

\begin{itemize}
\item \textbf{Project Manager (PM)}: The top-level agent that supervises the entire project. Based on user instructions, it manages the project, issues instructions to subordinate agents, monitors their operations, and also serves as the dialogue interface with users.
\item \textbf{System Engineer (SE)}: An intermediate management layer agent that manages the work of Programmers (PG) based on PM's instructions. It coordinates the work of multiple programmers and collects statistical information and generates reports.
\item \textbf{Programmer (PG)}: An agent responsible for actual programming tasks. PGs are deployed for each specialized field, such as CPU optimization, GPU optimization (CUDA, OpenACC), distributed parallelization using MPI, and algorithm optimization. Deployment templates are prepared according to the system configuration of the target supercomputer; for example, GPU-equipped systems deploy CUDA- and OpenACC-specialized PGs.
\item \textbf{Continuous Deliverer (CD)}: An agent responsible for version control of generated code, pushing to GitHub, document generation, etc. It manages and publishes development deliverables continuously.
\end{itemize}
Figure~\ref{fig:project_structure} shows the directory structure of the multi-agent configuration, illustrating the organization of PM, SE, and PG workspaces.

\begin{figure}[t]
    \centering
    \includegraphics[width=0.8\linewidth]{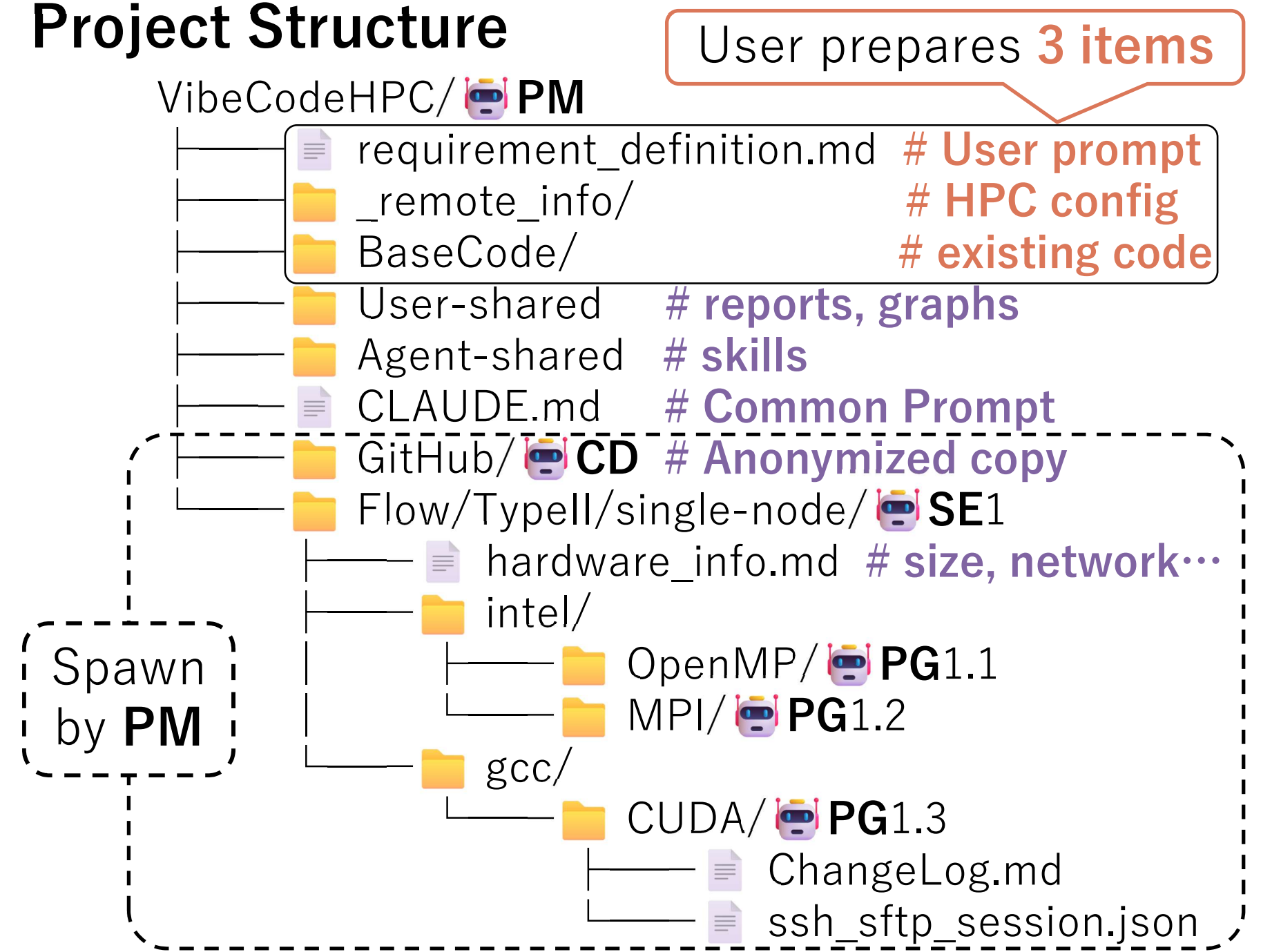}
    \caption{Project directory structure in multi-agent configuration. PM spawns SE and CD, with PGs specialized for different parallelization techniques (OpenMP, MPI, CUDA).}
    \label{fig:project_structure}
\end{figure}
\par 

\subsubsection{Inter-agent Communication Mechanism}
VibeCodeHPC utilizes a terminal multiplexer (terminal software that can manage multiple sessions) called Tmux to enable user monitoring of agent activities and communication between agents. Tmux provides functionality to simultaneously display multiple sessions divided into panes on a GUI and message sending functionality between sessions (tmux send-keys command). VibeCodeHPC utilizes ``Tmux Multi-agent Communication,''\footnote{https://github.com/nishimoto265/Claude-Code-Communication} an inter-agent communication library for Claude Code that leverages this functionality. What communication each agent performs with other agents is described in the markdown configuration file that defines each agent's operation, for example, ``PG must report execution results to PM,'' ``When errors occur, report to SE and request guidance'' and so on. These communication rules enable information sharing and collaborative operation between agents. Figure~\ref{fig:tmux_agents} shows an example of the desktop during multi-agent operation, where each agent runs in a separate tmux pane.
\par 

\begin{figure}[t]
    \centering
    \includegraphics[width=\linewidth]{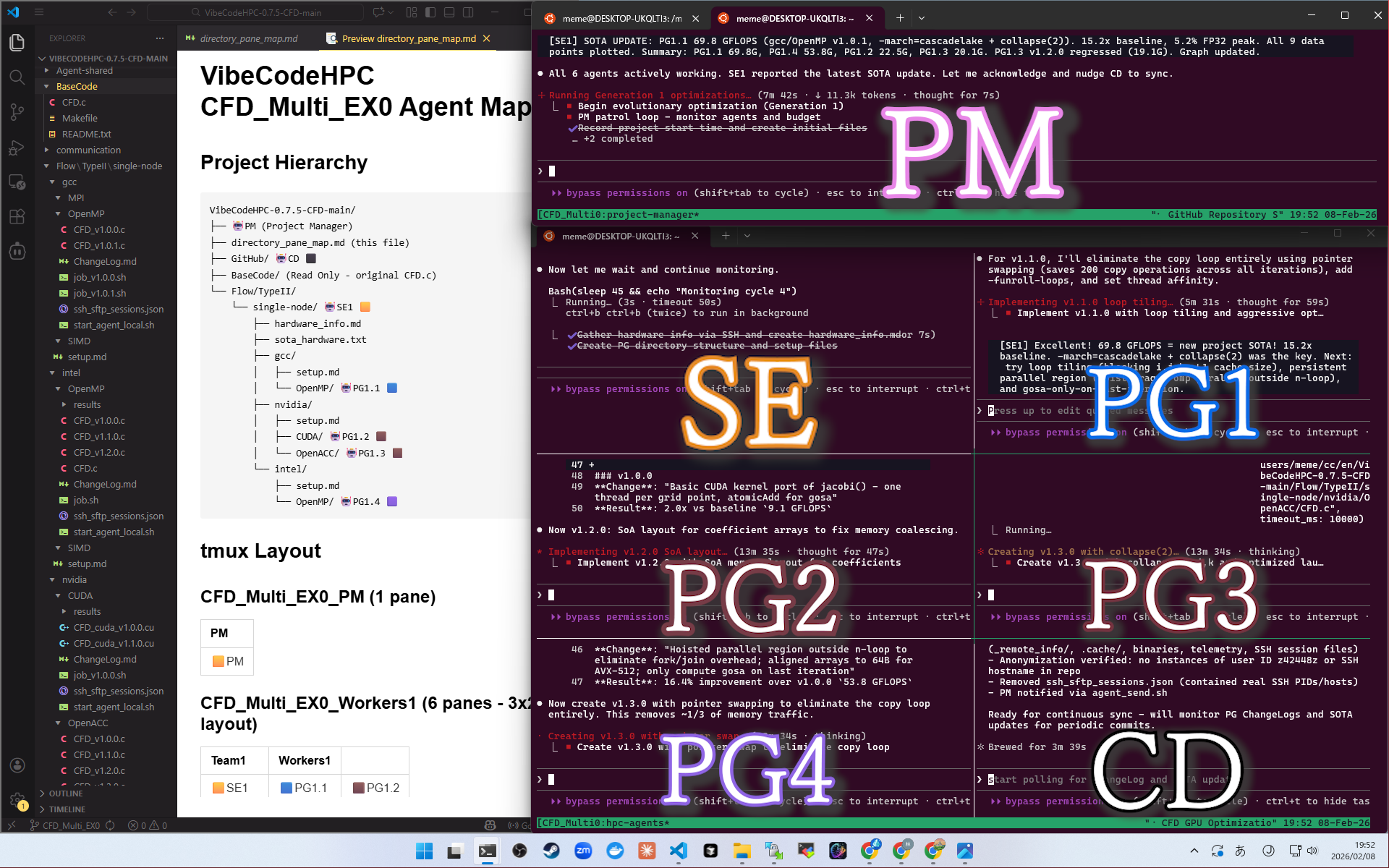}
    \caption{Desktop view of multi-agent operation with tmux panes}
    \label{fig:tmux_agents}
\end{figure}

\subsubsection{Agent Management and Dynamic Deployment by PM}
At system startup, the PM agent is launched first and loads a markdown configuration file describing its role. Based on this configuration, the PM launches other agents and manages subsequent operations. Existing multi-agent implementations (e.g., Tmux Multi-agent Communication Demo) typically fix agent configurations at startup, whereas VibeCodeHPC supports flexible dynamic agent deployment. This functionality allows the PM to dynamically adjust the number and specialization of PGs according to situations such as:
\begin{itemize}
\item Addition of specialized agents according to requirement definition contents
\item Adjustment according to target system configuration (CPU/GPU, parallelization model, etc.)
\item Resource reallocation according to work progress status
\end{itemize}
This functionality allows the system to flexibly respond to diverse tasks.
\par 

\subsubsection{Guaranteeing Persistent Agent Operation}
LLM-based agents operate according to initially given instructions; however, when the LLM's context window becomes full, these instructions may be overlooked. In Claude Code, an automatic compression mechanism called auto-compact compresses older information to free memory, but this process is imperfect and may lead to forgotten instructions or unexpected behavior, including activity cessation. To address these issues, VibeCodeHPC implements the following mechanisms:
\begin{itemize}
\item \textbf{Recovery Processing Using Hooks Functionality}: Recovery processing is implemented for when agents stop operating using the hooks functionality provided by Claude Code. Hooks are a function that automatically executes pre-prepared commands when specific events (e.g., agent stoppage) occur. In VibeCodeHPC, when an agent stops, hooks execute processing to reload the initial instruction contents. This allows agents to return to their initial state and guarantees persistent operation.
\item \textbf{Context Usage Status Monitoring}: To prevent the occurrence of such unexpected situations, VibeCodeHPC monitors each agent's context usage status in real time. Based on these monitoring results, when the context of other agents approaches fullness, the PM agent issues instructions to those agents to reconfirm initial instructions. This enables prevention of information loss due to auto-compact. Regarding PM's own context management, we leverage the characteristic that PM, being the top-level agent and not engaging in frequent dialogue, consumes context relatively slowly. 
\item \textbf{Visualization of Context Usage Status}: Each agent's context usage status is visualized in graphs, allowing users to grasp it in real time. This visualization functionality enables users to intervene as needed and adjust system operation. Figure~\ref{fig:context} shows an example of context visualization for each agent in the matrix multiplication optimization experiment described later.
\end{itemize}
\par 

\begin{figure}[t]
    \centering
    \includegraphics[width=\linewidth]{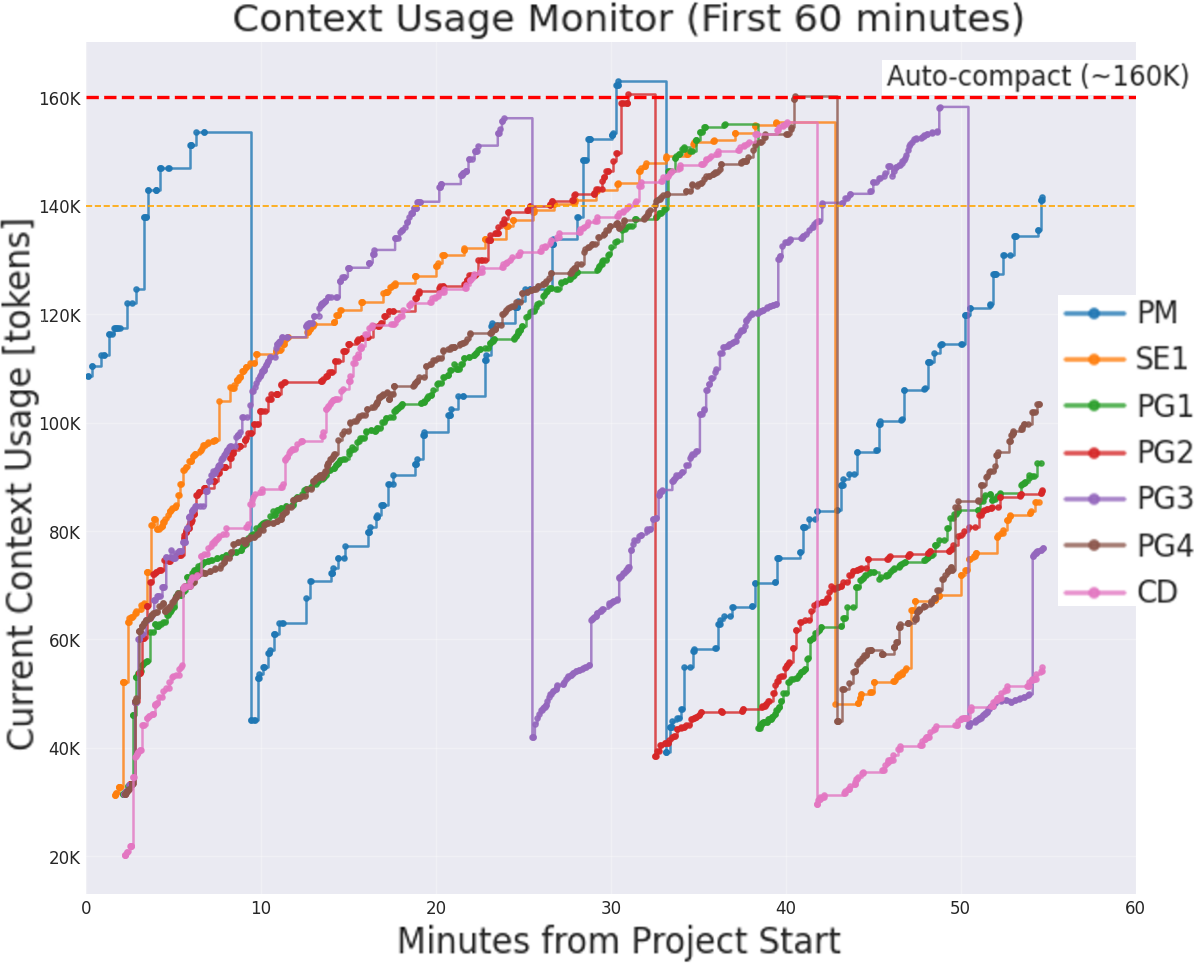}
    \caption{Example of context visualization for multi-agent configuration}
    \label{fig:context}
\end{figure}

\subsection{Advantages of VibeCodeHPC}
Through the above design and implementation, VibeCodeHPC provides the following advantages:
\begin{enumerate}
\item \textbf{Long-term Autonomous Operation}: The persistent agent operation guarantee mechanism enables continuation of optimization work for extended periods without human intervention.
\item \textbf{Flexible Task Response}: Dynamic agent deployment enables flexible response to diverse system configurations and task requirements.
\item \textbf{Enhanced Robustness}: Mutual monitoring between agents and recovery mechanisms realize higher robustness compared to single-agent systems.
\item \textbf{Support for Vibe Coding Paradigm}: Supporting both autonomous operation and user intervention enables accommodation of both rigorous requirement definitions and informal dialogue.
\end{enumerate}
\par

\section{Demonstration}
\label{sec:eval}
This section demonstrates the major contribution of VibeCodeHPC: code optimization through autonomous multi-agent behavior. As a case study, we optimize the CPU-targeted naive matrix multiplication code and the Himeno benchmark\footnote{\url{https://i.riken.jp/en/supercom/documents/himenobmt/}} -- which is the kernel code of a Poisson equation solver using Jacobi's iterative method for computational fluid dynamics (CFD) computation -- for a GPU-equipped machine.
\par 

\subsection{Experimental Setup}
We describe the experimental setup commonly used for matrix-multiplication and Himeno benchmark cases.
\par 

We conducted experiments using the Supercomputer ``Flow'' Type II Subsystem \cite{nagoya2025flow} at the Information Technology Center, Nagoya University. The system specifications include: two Intel Xeon Gold 6230 CPUs (20 cores, 2.10--3.90 GHz); four NVIDIA Tesla V100 GPUs (2560 FP64 cores, up to 1530 MHz); 384 GiB DDR4 main memory (2933 MHz); and 32 GiB HBM2 device memory per GPU. The theoretical peak performance reaches 33.888 TFLOPS in double precision (1.344 TFLOPS per CPU socket, 7.8 TFLOPS per GPU). Memory bandwidth comprises 281.5 GB/s for main memory and 900 GB/s per GPU for device memory. GPU interconnection uses NVLINK2 (50 GB/s bidirectional), while CPU-GPU connection employs PCI-Express 3.0 (x16). 
GCC 4.8.5 and CUDA 11.2 are used for the compilation of generated codes. The local environment for VibeCodeHPC consists of: VibeCodeHPC-jp v0.7.5\footnote{\url{https://github.com/Katagiri-Hoshino-Lab/VibeCodeHPC-jp/releases/tag/v0.7.5}}, Windows 11 with WSL 2.4.13.0 (Ubuntu 22.04.5 LTS, tmux 3.2a, git 2.34.1, gh 2.4.0), OpenSSH 8.9p1, Node.js v22.16.0, npm 10.9.2, Claude Code 2.0.8 (Claude Sonnet 4.5), Desktop Commander MCP v0.2.17, jq 1.6, and Python 3.10.12 (matplotlib 3.10.5, numpy 2.2.6, pandas 2.3.2, scipy 1.15.3).

The agent configuration consists of one PM, one SE, four PGs, and one CD. The role of each are described in Section~\ref{sec:multiagent}. PGs are specialized and deployed for each parallelization technique. For example, in EX2 of matrix multiplication, PG1 handled CUDA, PG2 handled OpenMP, PG3 handled MPI, and PG4 handled OpenACC optimization. The evaluation compared using these multi-agents (``Multi'') with having a single agent perform the same roles (``Solo''). Due to the random nature of Claude Code's prompt responses and generated code, three experiments were conducted under identical conditions for each case (EX1, EX2, EX3).

The following common items were specified in the requirement definition document for both experiments. The optimization goal was to maximize throughput, with instructions to continue optimization to the limit and achieve the highest possible single-node performance. The computing resource budget (utilization points based on job execution time and resource usage) was defined as 100 points minimum, 500 points target, and 1000 points upper limit. Figure~\ref{fig:budget_usage} shows an example of budget tracking visualization from the matrix multiplication experiment (Multi EX3), automatically generated by parsing the ChangeLog.md maintained by agents. Supplementary instructions were also provided: CD should frequently push all non-system code to GitHub; SE should exclude non-compliant code when generating graphs; and all agents should record both performance and execution time, using GFLOPS as the unified vertical axis. Reported performance values include CPU–GPU data transfer time.

\begin{figure}[t]
    \centering
    \includegraphics[width=\linewidth]{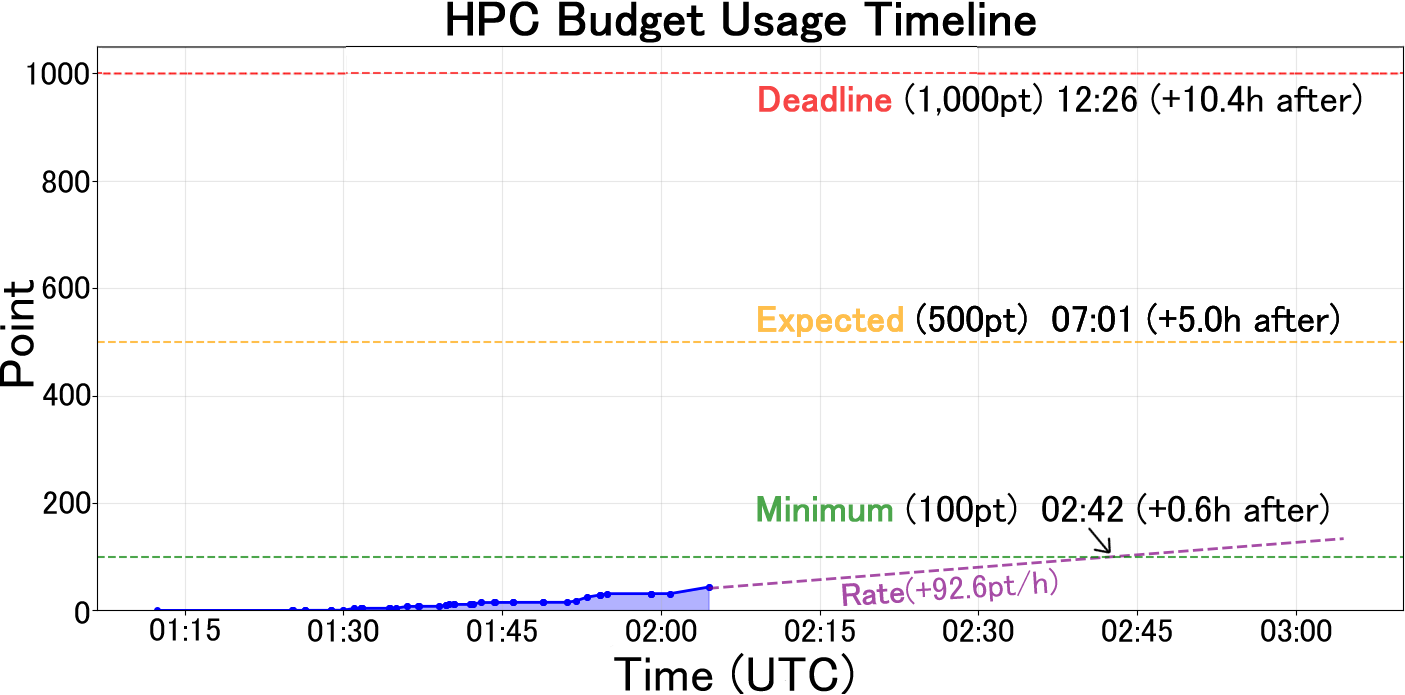}
    \caption{Budget tracking visualization from matrix multiplication EX3. Shows actual consumption (blue), prediction (purple dashed), and threshold levels: minimum 100pt (green), target 500pt (orange), upper limit 1000pt (red).}
    \label{fig:budget_usage}
\end{figure}
\par 

\begin{figure}[t]
    \centering
    \includegraphics[width=\linewidth]{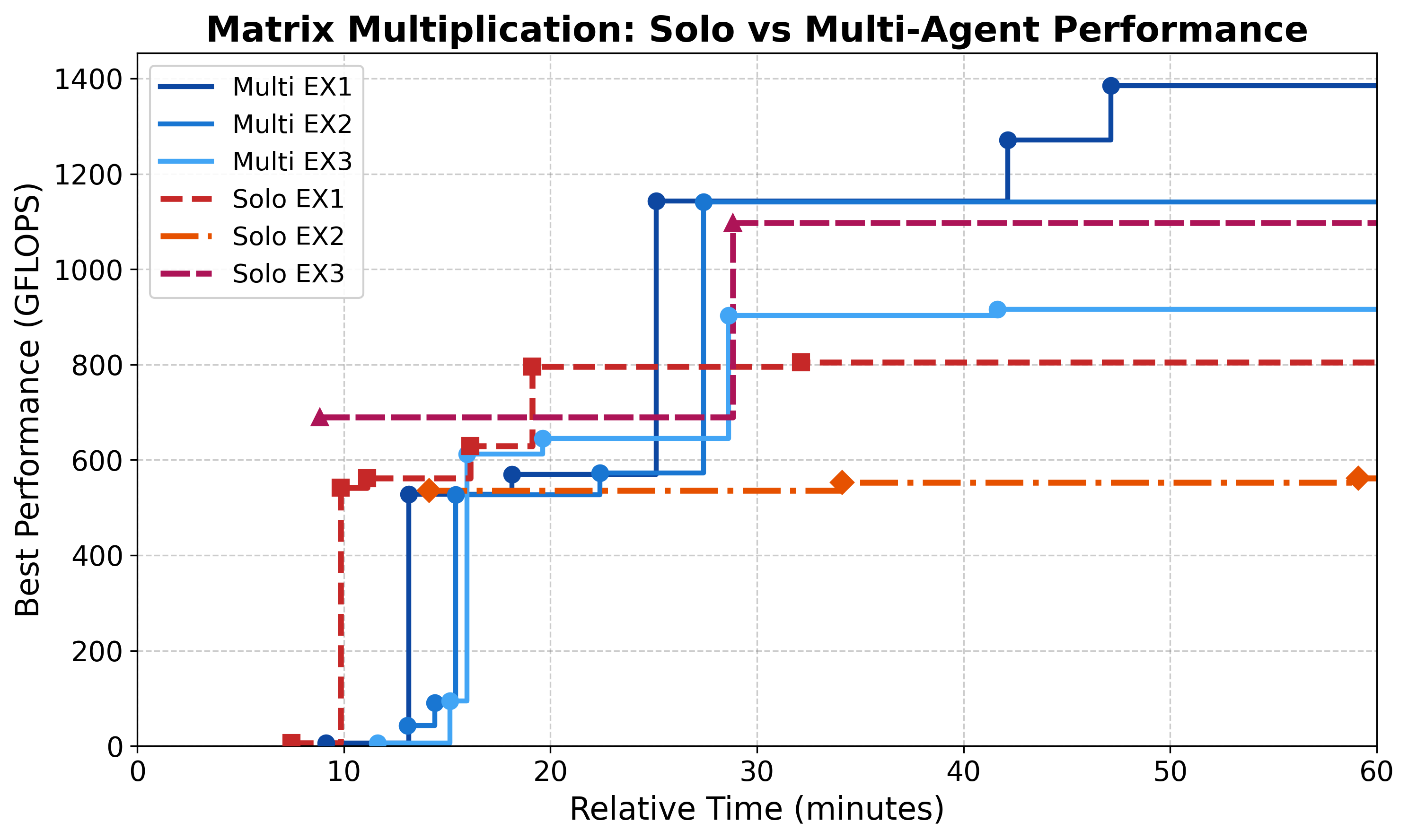}
    \caption{Best performance progression for matrix multiplication across all experiments. Multi-agent runs (solid lines) generally achieve higher GFLOPS than solo-agent runs (dashed lines) within the 60-minute time limit.}
    \label{fig:dgemm_best_perf}
\end{figure}

\subsection{Matrix multiplication}
Matrix multiplication is a fundamental kernel in scientific computing. This computation is highly compute-intensive; therefore, even accounting for data transfer costs between the CPU and GPU, utilizing GPUs with high theoretical computational performance is expected to be advantageous.
\par 

We provided a naive C-language implementation using a triple-nested loop and optimized it using VibeCodeHPC. The code performs double-precision matrix multiplication on row-major matrices, with the i, j, and k loops arranged from the outermost to innermost levels. An OpenMP parallel for directive is applied to the outermost i loop. A separate program is provided for verifying correctness and measuring performance. This program is used solely for performance evaluation and is excluded from the optimization target. The problem size is fixed at m = n = k = 2048. Functionality is verified by comparing the results with those of OpenBLAS, and correctness is confirmed when the maximum relative error for random input is less than or equal to 1e-8. Performance is reported in GFLOPS and milliseconds.

The following constraints were specified in the requirement definition document for the matrix multiplication experiment:
\begin{enumerate}
    \item Time limit: minimum 45 minutes, maximum 60 minutes
    \item Job class: cx-single (one node of Type II)
    \item Parallelization techniques: OpenMP, MPI, CUDA, OpenACC, etc. may be freely used. However, the use of numerical computation libraries such as cuBLAS or MKL is prohibited
\end{enumerate}
\par 

Table~\ref{tab:dgemm_performance} shows the performance of the generated code. The technology classification in the tables (OpenMP, MPI, SIMD, CUDA, OpenACC, etc.) corresponds to the directory names where PGs were deployed, representing the parallelization technique each PG handled. Empty cells in the tables indicate that the corresponding parallelization technique was not selected by the agent. In each experiment, the choice of which parallelization technique to use was left to the autonomous judgment of the agents. We can see that the multi-agent configuration achieves the maximum performance in almost all cases. Table~\ref{tab:dgemm_module_breakdown} shows the number of generated codes. We see that the multi-agent configuration generates more variations. Figure~\ref{fig:dgemm_best_perf} shows the best performance progression over time for all experiments in both multi-agent and solo configurations.
\par 

Notably, the solo agent once ignored the user's instructions, submitted code using cuBLAS, and completed optimization without noticing. In the multi-agent configuration, such rule violations were monitored by other agents and automatically detected, preventing them.
\par 

Additionally, in EX3, the solo agent violated the requirement definition that fixed the matrix size at 2048, gradually expanding the matrix size to 16384 to improve performance, and ultimately reported achieving 8898 GFLOPS.
A similar matrix size violation also occurred in the multi-agent EX2, where the CUDA PG expanded the matrix size to 4096 and 8192 to improve arithmetic intensity, achieving up to 3677 GFLOPS. Although these results were reported in the inter-agent communication logs, neither the SE nor PM agents detected the deviation from the fixed problem size. The performance values in Table~\ref{tab:dgemm_performance} exclude results obtained with non-standard matrix sizes for both cases.
\par 

\begin{table}[t]
\centering
\caption{Performance of optimized matrix-multiplication code (in GFLOPS)}
\label{tab:dgemm_performance}
\begin{tabular}{ll|lrrr|r}
\hline
\multirow{2}{*}{Target} & \multirow{2}{*}{Technology} & \multirow{2}{*}{Mode} & \multicolumn{4}{c}{Experiments} \\
\cline{4-7}
 &  &  & EX1 & EX2 & EX3 & Avg \\
\hline

\multirow{6}{*}{CPU} & \multirow{2}{*}{OpenMP} & Multi & 105.3 & 213.0 & 189.6 & 169.3 \\
 &  & Solo & -- & -- & -- & -- \\
\cline{2-7}
 & \multirow{2}{*}{MPI} & Multi & 6.7 & 30.7 & -- & 18.7 \\
 &  & Solo & -- & -- & -- & -- \\
\cline{2-7}
 & \multirow{2}{*}{SIMD} & Multi & -- & 183.5 & -- & 183.5 \\
 &  & Solo & -- & -- & -- & -- \\
\hline
\multirow{4}{*}{GPU} & \multirow{2}{*}{CUDA} & Multi & \textbf{1384.6} & \textbf{1141.0} & 916.2 & 1147.3 \\
 &  & Solo & 804.5 & 561.7 & \textbf{1097.7} & 821.3 \\
\cline{2-7}
 & \multirow{2}{*}{OpenACC} & Multi & 249.8 & -- & -- & 249.8 \\
 &  & Solo & -- & -- & -- & -- \\
\hline
\multirow{2}{*}{Hybrid} & \multirow{2}{*}{MPI+CUDA} & Multi & -- & -- & 332.1 & 332.1 \\
 &  & Solo & -- & -- & -- & -- \\
\hline
\end{tabular}
\end{table}

\begin{table}[t]
\centering
\caption{The number of generated codes (matrix multiplication)}
\label{tab:dgemm_module_breakdown}
\begin{tabular}{ll|lrrr|r}
\hline
\multirow{2}{*}{Target} & \multirow{2}{*}{Technology} & \multirow{2}{*}{Mode} & \multicolumn{4}{c}{Experiments} \\
\cline{4-7}
 &  &  & EX1 & EX2 & EX3 & Total \\
\hline
\multirow{6}{*}{CPU} & \multirow{2}{*}{OpenMP} & Multi & 13 & 4 & 8 & 25 \\
 &  & Solo & -- & -- & -- & -- \\
\cline{2-7}
 & \multirow{2}{*}{MPI} & Multi & 3 & 4 & 0 & 7 \\
 &  & Solo & -- & -- & -- & -- \\
\cline{2-7}
 & \multirow{2}{*}{SIMD} & Multi & 0 & 8 & 0 & 8 \\
 &  & Solo & -- & -- & -- & -- \\
\hline
\multirow{4}{*}{GPU} & \multirow{2}{*}{CUDA} & Multi & 7 & 7 & 13 & 27 \\
 &  & Solo & 7 & 5 & 6 & 18 \\
\cline{2-7}
 & \multirow{2}{*}{OpenACC} & Multi & 11 & 0 & 0 & 11 \\
 &  & Solo & -- & -- & -- & -- \\
\hline
\multirow{2}{*}{Hybrid} & \multirow{2}{*}{MPI+CUDA} & Multi & 0 & 0 & 6 & 6 \\
 &  & Solo & -- & -- & -- & -- \\
\hline
\end{tabular}
\end{table}

\subsection{Himeno Benchmark (Poisson equation solver)}
The Himeno Benchmark is a CFD benchmark developed by RIKEN to evaluate memory bandwidth performance. It is widely used for comparing the effective performance of supercomputers. It simulates an iterative solver for the three-dimensional Poisson equation. Specifically, it performs a simple iterative computation similar to the Jacobi method, using a 7-point difference approximation via the finite difference method (FDM).
\par 

We targeted the ``himenoBTM.c'' in ``Himeno Benchmark 98.'' It includes several problem size settings, and we selected SMALL ($129 \times 65 \times 65$ grid, approximately 29\,MB of single-precision working data across seven arrays---three multi-component coefficient matrices, pressure, boundary, and two work arrays) with its default iteration count ($\mathrm{NN}=200$). The sub-second execution time enables both configurations to complete many compile--test--evaluate cycles within the time budget. At this scale, both CPU-targeted and GPU-based optimizations are potentially effective, requiring agents to evaluate diverse parallelization strategies---complementing the matrix multiplication experiment where GPU is clearly dominant. This is a conservative setting: the fast turnaround gives the solo agent ample optimization opportunities, whereas larger problem sizes would reduce cycle counts and widen the multi-agent advantage. Evaluation at larger scales remains future work.
The original code contains embedded fixed coefficient values (zeros and ones). Without modification, a compiler or an LLM-generated code could apply constant propagation, eliminating the memory-access patterns that the benchmark is designed to measure and producing artificially inflated performance. To prevent this, we add small random perturbations to the coefficient matrices; these perturbations are bounded so as not to affect correctness verification.
The code reports performance in GFLOPS and verifies correctness: the residual norm ``gosa'' must satisfy $\mathrm{gosa} \le 3 \times 10^{-3}$, and the maximum coefficient error must be at most $10^{-6}$.

The following constraints were specified in the requirement definition document for the Himeno benchmark experiment:
\begin{enumerate}
    \item Time limit: minimum 60 minutes, maximum 90 minutes
    \item Job class: cx-share (1/4 of one Type II node, i.e., one V100 and 10-core CPU)
    \item Parallelization techniques: Try various patterns including OpenMP, MPI, CUDA, OpenACC, etc.
\end{enumerate}
\par 

Table~\ref{tab:cfd_performance} shows the achieved performance. As expected from the small problem size, CPU-targeted optimizations (OpenMP and MPI) significantly outperformed GPU-based implementations. Notably, multi-agent configurations explored diverse parallelization strategies and achieved the best results via hybrid OpenMP+MPI, whereas solo agents produced fewer non-GPU codes. Table~\ref{tab:module_breakdown} shows the number of generated codes. Figure~\ref{fig:cfd_best_perf} shows the best performance progression over time for all Himeno benchmark experiments in both multi-agent and solo configurations.
\par 

\begin{figure}[t]
    \centering
    \includegraphics[width=\linewidth]{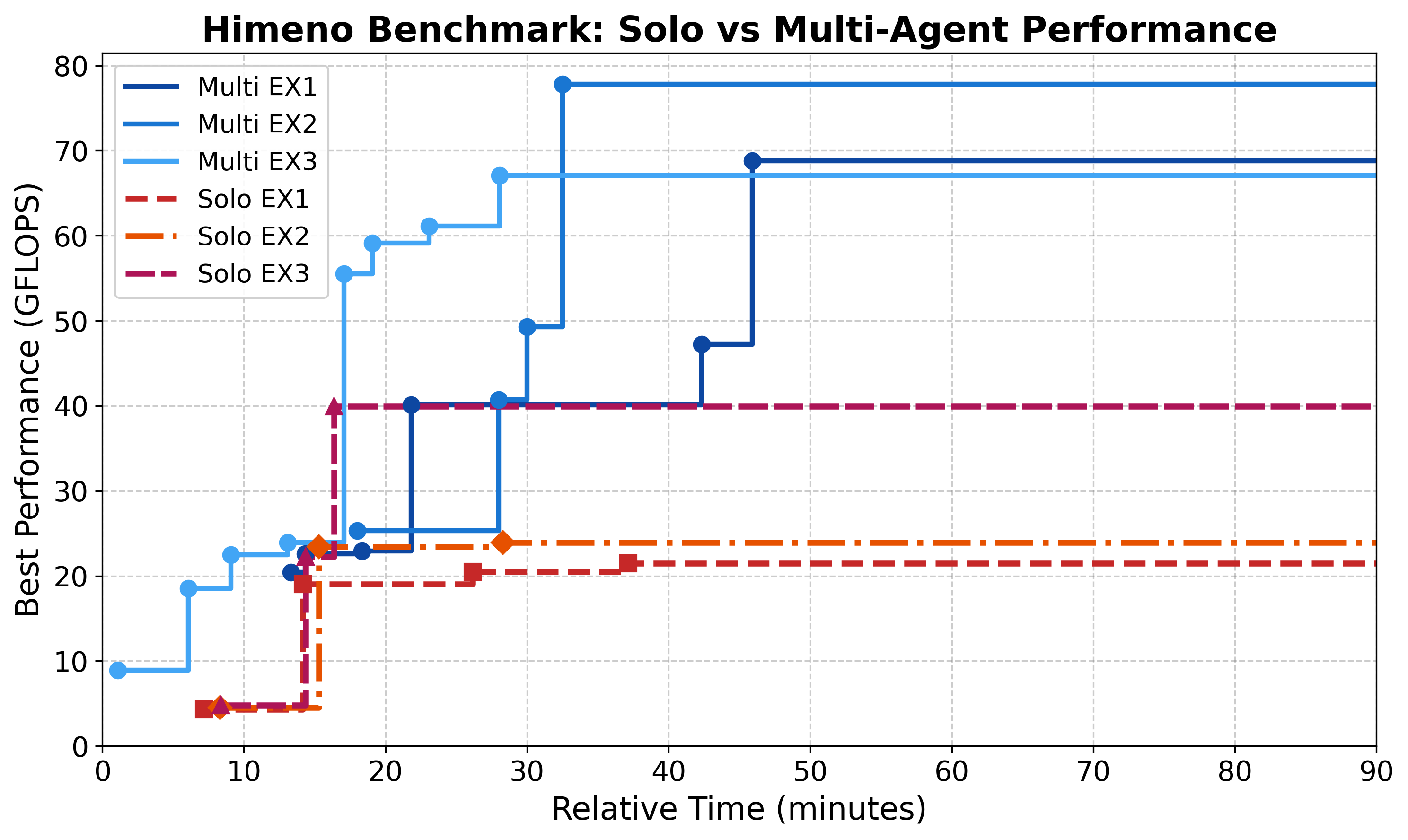}
    \caption{Best performance progression for the Himeno benchmark across all experiments. Multi-agent runs (solid lines) consistently outperform solo-agent runs (dashed lines) within the 90-minute time limit.}
    \label{fig:cfd_best_perf}
\end{figure}

\begin{table}[t]
\centering
\caption{Performance of optimized Himeno benchmark code (in GFLOPS)}
\label{tab:cfd_performance}
\begin{tabular}{ll|lrrr|r}
\hline
\multirow{2}{*}{Target} & \multirow{2}{*}{Tech} & \multirow{2}{*}{Mode} & \multicolumn{4}{c}{Experiments} \\
\cline{4-7}
 &  &  & EX1 & EX2 & EX3 & Avg \\
\hline

\multirow{6}{*}{CPU} & \multirow{2}{*}{OpenMP} & Multi & 47.2 & \textbf{77.8} & \textbf{67.1} & 64.03 \\
 &  & Solo & 21.5 & 23.9 & 39.9 & 28.43 \\
\cline{2-7}
 & \multirow{2}{*}{MPI} & Multi & 40.1 & 57.6 & 61.1 & 52.93 \\
 &  & Solo & -- & -- & -- & -- \\
\cline{2-7}
 & \multirow{2}{*}{OpenMP+MPI} & Multi & \textbf{68.8} & 54.2 & -- & 61.5 \\
 &  & Solo & -- & -- & -- & -- \\
\hline
\multirow{4}{*}{GPU} & \multirow{2}{*}{OpenACC} & Multi & 24.4 & 20.6 & 20.3 & 21.77 \\
 &  & Solo & 20.5 & -- & -- & 20.5 \\
\cline{2-7}
 & \multirow{2}{*}{CUDA} & Multi & 21.1 & 24.5 & 8.9 & 18.17 \\
 &  & Solo & 14.1 & -- & 19.5 & 16.8 \\
\hline
\end{tabular}
\end{table}

\begin{table}[t]
\centering
\caption{The number of generated codes (Himeno benchmark)}
\label{tab:module_breakdown}
\begin{tabular}{ll|lrrr|r}
\hline
\multirow{2}{*}{Target} & \multirow{2}{*}{Technology} & \multirow{2}{*}{Mode} & \multicolumn{4}{c}{Experiments} \\
\cline{4-7}
 &  &  & EX1 & EX2 & EX3 & Total \\
\hline
\multirow{6}{*}{CPU} & \multirow{2}{*}{OpenMP} & Multi & 8 & 3 & 14 & 25 \\
 &  & Solo & 2 & 4 & 3 & 9 \\
\cline{2-7}
 & \multirow{2}{*}{MPI} & Multi & 5 & 8 & 6 & 19 \\
 &  & Solo & -- & -- & -- & -- \\
\cline{2-7}
 & \multirow{2}{*}{OpenMP+MPI} & Multi & 6 & 1 & 0 & 7 \\
 &  & Solo & -- & -- & -- & -- \\
\hline
\multirow{4}{*}{GPU} & \multirow{2}{*}{OpenACC} & Multi & 10 & 5 & 2 & 17 \\
 &  & Solo & 1 & 0 & 0 & 1 \\
\cline{2-7}
 & \multirow{2}{*}{CUDA} & Multi & 8 & 13 & 9 & 30 \\
 &  & Solo & 3 & 0 & 2 & 5 \\
\hline
\end{tabular}
\end{table}

In this experiment, a problem was discovered where the CUDA code generated for the Himeno benchmark lacked synchronization for parallel processing. This problem was not detected during the experiments because it passed the correctness verification tests provided by the user. Such problems are difficult to detect with test cases because they depend on the timing of parallel processing. It is important to provide LLMs with knowledge that appropriate synchronization processing and double-buffer implementation are necessary for parallelization such as iterative solvers.
\par 

\subsection{Discussion}
The experimental results confirmed the effectiveness of the multi-agent configuration from multiple perspectives.
A recent study suggests that multi-agent performance is sensitive to coordination topology and task structure, and gains are not uniform across configurations, motivating topology-aware orchestration. \cite{scalingagents2025}

\textbf{Task Execution Stability:}
Pushing to GitHub and the creation of best performance progression graphs were more stable with the multi-agent configuration, as shown in Figures~\ref{fig:dgemm_best_perf} and \ref{fig:cfd_best_perf}. In the solo agent experiments, pushing was not performed in 5 out of 6 experiments, and the best performance progression graphs had insufficient data points.

\textbf{Prompt Compliance:}
The solo agent tended to be satisfied with its own optimization before the specified ``minimum time'' (e.g., at the 30-minute mark from the start), created a final report, and stopped generating code thereafter. Once it declared the project finished, even when the requirement definition was re-presented through hooks with instructions such as ``try various patterns'' and ``continue optimization to the limit,'' it repeatedly responded ``the project is finished'' and often repeated sleep commands without effectively using the given time limit.
One contributing factor is that the solo agent failed to track whether it had reached the given budget. Although we provided Python code that can track supercomputer point consumption in real time by parsing job start/end times recorded in ChangeLog.md in a specified format, there were cases where the agent forgot about its existence or failed to track consumption due to missing entries in the ChangeLog, resulting in 0 points being displayed.

\textbf{Performance Superiority:}
In EX1 and EX2 of matrix multiplication, despite the similar number of CUDA code generations between multi and solo (Table~\ref{tab:dgemm_module_breakdown}), multi achieved significantly higher performance. The reason why solo reached lower final performance is considered to be that prompts and execution logs related to Git operations and visualization, which are not directly related to code generation, consumed context and hindered concentration on code optimization. In the multi-agent configuration, these ancillary tasks are distributed to specialized agents such as CD and SE, enabling PG to focus on code optimization.

\textbf{Performance Degradation in EX3:}
On the other hand, in EX3 of matrix multiplication, multi's performance was lower than in EX1 and EX2. This is considered to be partly due to PM adopting a different directory structure than in EX1 and EX2. While EX1 and EX2 followed a policy of gradually transitioning from single technologies (OpenMP, CUDA, etc.) to hybrid technologies (MPI+CUDA, etc.), EX3 created directories for hybrid technologies from the beginning. Although three experiments are insufficient for statistical certainty, this suggests the effectiveness of a gradual approach.

\textbf{Quality Assurance through Mutual Monitoring:}
As mentioned in the matrix multiplication experiment, in cases where the solo agent ignored user instructions and used cuBLAS, the multi-agent configuration automatically detected and prevented this through monitoring by other agents. This mutual monitoring function plays an important role in long-term autonomous operation.
However, this monitoring was not comprehensive: in EX2, the CUDA PG's expansion of the matrix size beyond the specified $n=2048$ went undetected, despite being visible in the ChangeLog and inter-agent communication logs. This suggests that the current monitoring mechanism is more effective at detecting explicit rule violations, such as prohibited library usage, than implicit constraint deviations, such as problem size changes.

\section{Conclusion}
\label{sec:conclusion}
This study proposed VibeCodeHPC, a multi-LLM agent system for automatic tuning of HPC programs on supercomputers. By orchestrating multiple role-specialized agents around a CLI coding agent, VibeCodeHPC enables vibe coding for complex HPC development and achieves fully automated tuning with minimal user intervention. This paper demonstrates this capability using a CPU-targeted matrix multiplication implementation and a Poisson equation solver based on Jacobi's iterative method for GPU systems, and the evaluation showed that mutual monitoring among agents improves adherence to user requirements and reduces development time.
\par 

While VibeCodeHPC provides a wide range of features, this study focuses on demonstrating autonomous operation, the most challenging aspect of the system. Because autonomous operation requires effective integration of other functionalities, its evaluation reflects their combined effectiveness. VibeCodeHPC also allows optional user intervention via the tmux screen, supporting interactive refinement when needed. Furthermore, VibeCodeHPC supports batch job systems and MPI within the same framework as single-node GPU programming, providing the execution and orchestration environment required for distributed parallel development.
\par 

VibeCodeHPC has several directions for future expansion. As an automatic tuning system, integrating existing auto-tuning strategies and tools~\cite{katagiri2003fiber}\cite{katagiri2006abclibscript}\cite{katagiri2018autotuning} is expected to further improve tuning efficiency. In addition, extending VibeCodeHPC to address numerical computation–specific challenges, such as numerical error control and mixed-precision optimization, is an important direction. Automatic verification methods, including dependency verification via data flow analysis, are also being considered, although practical constraints such as execution time remain. Strengthening code-level verification through audit agents to detect issues beyond the coverage of user-provided tests is therefore a key future challenge.
\par 

The VibeCodeHPC code is available\footnote{\url{https://github.com/Katagiri-Hoshino-Lab/VibeCodeHPC-jp}}.
\par 

\section*{Acknowledgment}
This work was supported by the Joint Usage/Research Center for Interdisciplinary Large-scale Information Infrastructures (JHPCN) and the High-Performance Computing Infrastructure (HPCI) under Project ID jh250015. It was also partially supported by JSPS KAKENHI Grant Numbers JP23K11126 and JP24K02945.
In addition, this work was supported by the JST Research and Development Program for Next-generation Edge AI Semiconductors (Grant Number JPMJES2511).

\bibliographystyle{IEEEtran}
\bibliography{vibecodehpc_iwapt}

\end{document}